\documentclass[]{raa}
\usepackage{deluxetable}
\usepackage{amssymb}            % referee version: for submission
\usepackage{graphicx,times}
\usepackage{natbib}
\usepackage{color}

\providecommand{\bm}[1]{\mbox{\bol
dmath$#1$\unboldmath}}

\begin{document}
%\SetRunningHead{H.-L. Li
%ET AL.}{THE SOLAR-TYPE BINARY OO AQUILAE}
%\Received{~~~~/~~/~~}%{yyyy/mm/dd}
%\Accepted{~~~~/~~/~~}%{yyyy/mm/dd}

\title{B3\,0749+460A: A New ``Changing-Look'' Active Galactic Nucleus Associated with X-ray Spectral Slope Variations}

% \volnopage{ {\bf 2009} Vol.\ {\bf 9} No. {\bf 10}, 1103--1118}
\volnopage{}
	\setcounter{page}{1}

	\author{J. Wang\inst{1,2} \and W. K. Zheng\inst{3} \and D. W. Xu\inst{2,4} \and T. G. Brink\inst{3} \and 
	A. V. Filippenko\inst{3,5} \and  C. Gao\inst{1,2,4} \and S. S. Sun\inst{1,2,4} \and J. Y. Wei\inst{2,4}
	}
 \institute{Guangxi Key Laboratory for Relativistic Astrophysics, School of Physical Science and Technology, Guangxi University, Nanning 530004, China; {wj@nao.cas.cn} 
  \and Key Laboratory of Space Astronomy and Technology, National Astronomical Observatories,
Chinese Academy of Sciences, Beijing 100101, China; {dwxu@nao.cas.cn}
  \and Department of Astronomy, University of California, Berkeley, CA 94720-3411, USA
  \and School of Astronomy and Space Science, University of Chinese Academy of Sciences, Beijing, People's Republic of China
  \and Miller Institute for Basic Research in Science, University of California, Berkeley, CA 94720, USA\\
%	University of Chinese Academy of Sciences,
%Beijing 100049, China; \\
%   \and
%	School of Physics and Electronic Information, Huaibei
%Normal University,235000 Huaibei, Anhui Province, China;\\
%   \and
%	School of Physics and Electronic Information, Huaibei
%Normal University,235000 Huaibei, Anhui Province, China;
%
%\vs \no
%	{\small Received * ; accepted *}
}

\abstract{
Here we report an identification of B3\,0749+460A as a new double-peaked 
local  ``changing-look'' active galactic nucleus (CL-AGN) in terms of our multi-epoch spectroscopic analysis. By comparing our new spectra taken in 2021
with the ones taken by the Sloan Digital Sky Survey in 2004, BOSS in 2013, and MaNGA in 2016, we reveal type transitions of Seyfert (Sy) 1.9$\rightarrow$Sy1.8$\rightarrow$Sy1.9. 
In the transitions,  the classical broad H$\alpha$ emission fades away since 2013, and 
disappears in our 2021 spectrum, although the absence of
broad H$\beta$ can be traced back to at least 2016.
A follow-up observation in X-rays by the {\it Swift}/XRT reveals that (1)
the X-ray emission level gradually decreases since 2005; and (2) the X-ray spectrum is soft in the optical ``turn-off'' state and hard in the `turn-on'' state. We argue that the disappearance of
the classical broad H$\alpha$ emission can be likely explained by the 
disk-wind broad-line-region model, in which the CL phenomenon is sensitive to luminosity in individual AGNs.
}
%\KeyWords{ binaries: close --- binaries: eclipsing --- stars: individual (OO Aql)} }

 \authorrunning{Wang et al. }            %author_head in even pages
 \titlerunning{X-ray Spectral Slope in CL-AGN B3\,0749+460}  % title_head in odd pages
 \maketitle

\section{Introduction}

Depending on the existence of broad Balmer emission lines (full width at
half-maximum intensity (FWHM) $\gtrsim 2000$\,km\,s$^{-1}$) 
in their optical spectra, active galactic nuclei (AGNs) can be classified into 
Type~1 and Type~2. 
The latter lack broad Balmer emission lines in their spectra, widely understood by the unified model (e.g., Antonucci 1993) in which its central engine is 
obscured by the dusty torus along the line of sight to an observer.
This orientation-based model has, however, recently been 
challenged by the rarely
discovered ``changing-look'' (CL) phenomenon, in which some AGNs show a spectral transition between Type~1, intermediate, 
and Type~2 within a timescale of years to decades (e.g., MacLeod et al. 2010, 2016; Shapovalova et al. 2010; Shappee et al. 2014;
LaMassa et al. 2015; McElroy et al. 2016; Ruan et al. 2016;
Runnoe et al. 2016; Parker et al. 2016; Gezari et al. 2017; Sheng et al. 2017, 2020; Stern et al. 2018; Yang et al. 2018; Wang et al. 2018b, 2019, 2020a; Frederick et al. 2019; Trakhtenbrot et al. 2019; Yan et al. 2019;  Guo et al. 2019; Ai et al. 2020; Graham et al. 2020; Kollatschny et al. 2018, 2020). 
The widely accepted standard disk model has also been challenged by the 
CL phenomenon in terms of the viscosity crisis (e.g., Lawrence 2018, and references therein).

The physical origin of CL-AGNs is still an open question, although the scenario involving clumpy 
obscuration can be almost entirely excluded by light echos in the mid-infrared  (e.g., Sheng et al. 2017) and by 
spectropolarimetry (e.g., Hutsemekers et al. 2019). 
There is, in fact, some evidence supporting the idea that the CL phenomenon results from a 
variation of the accretion power of a supermassive black hole (SMBH, e.g., H. Feng et al. 2021), even though the underlying physical connection 
is still poorly understood. On the one hand, some previous studies argue that CL-AGNs tend to be biased toward the low Eddington ratio end 
($L/L_{\mathrm{Edd}}; L_{\mathrm{Edd}} = 1.26\times10^{38} M_{\mathrm{BH}}/M_\odot\  \mathrm{erg\ s^{-1}}$ 
is the Eddington luminosity; e.g., MacLeod et al. 2019; Wang et al. 2020b), which is roughly consistent with the expectation of the  
disk-wind broad-line region (BLR) model (e.g., Elitzur \& Ho 2009; Nicastro 2000; Elitzur \& Shlosman 2006).

On the other hand, Ruan et al. (2019)  and Ai et al. (2020, and references therein) argue that the CL phenomenon
might be related to an
accretion-state transition similar to that occurring in X-ray binaries (XRBs), after taking into account 
a ``V''-shaped correlation between X-ray hardness and $L/L_{\mathrm{Edd}}$ identified in a few CL-AGNs.
This conclusion, however, 
is argued against by Wang et al. (2020b), who found that the CL phenomenon identified in UGC\,3223 is related to its X-ray emission level, rather than to its X-ray hardness ratio.

At the current stage, understanding of the CL phenomenon is greatly hampered by the
scarcity of identified cases; 
there are only $\sim100$ CL-AGNs identified by multi-epoch spectroscopy. Among these,
there are just 8 objects showing repeat type transitions
(e.g., Parker et al. 2019; Wang et al. 2020a;  Marin et al. 2019, and references therein), and only a couple of 
cases with a comparison study in X-rays.

Hon et al. (2020) recently reported a significant line-profile variation in the local AGN B3\,0749+460A.
In this paper, we identify the object as a new double-peaked CL-AGN thanks to our elaborate spectral analysis, and we report a follow-up observation in X-rays, which enables us to claim an X-ray spectral slope-dependent CL phenomenon.
The paper is organized as follows. 
Section 2 presents our optical spectroscopic and X-ray observations, along with the data reduction. 
The spectral analysis in both optical and X-rays is described in Section 3. Sections 4 and 5 present the results and discussion, respectively. 
A significant line-profile variation in this object was reported by Hon et al. (2020), who argue 
that it is an unusual object; we include a comparison with this study.  
A $\Lambda$CDM cosmological model with parameters H$_0=70\,\mathrm{km\,s^{-1}\,Mpc^{-1}}$, $\Omega_{\mathrm{m}}=0.3$, and 
$\Omega_\Lambda=0.7$ is adopted throughout.

\section{Observations and Data Reduction}

B3\,0749+460A (=NPM1G+46.0092, $\alpha = \mathrm{07^h52^m44^s.2}$, $\delta = \mathrm{+45\degr56\arcmin58\arcsec}$; J2000)
is a local AGN at a redshift of $z=0.0518$. The object was classified as a Seyfert 1.9 galaxy in the catalog of quasars and 
active nuclei (12th ed.; Veron-Cetty \& Veron 2006) and the NASA/IPAC Extragalactic Database (NED)\footnote{http://ned.ipac.caltech.edu/.}.
It is part of a sample of CL-AGN candidates we selected by cross-matching the local partially obscured AGNs
extracted from the Sloan Digital Sky Survey (SDSS) DR7 with the 
{\it Wide-field Infrared Survey Explorer (WISE)} catalog (Wright et al. 2010).

\subsection{Optical Spectroscopy}
\subsubsection{Observations}

We performed long-slit spectroscopic observations with the 2.16\,m telescope (Fan et al. 2016) at the Xinglong Observatory
of the National Astronomical Observatories, Chinese Academy of Sciences (NAOC) on 2021 January 12 (UT dates are used throughout this paper), and with
the 3\,m Shane telescope at Lick Observatory on  2021 January 21.
The first spectrum was taken with the Beijing Faint Object Spectrograph and
Camera (BFOSC) that is equipped with a back-illuminated E2V55-30 AIMO CCD.
We used a long slit of width 2\arcsec\ oriented in the north–south direction.
Our spectral resolution was $\sim10$\,\AA\ and the wavelength coverage was 3850--8200\,\AA.
The exposure time was $2 \times 2400$\,s.
Wavelength calibration was carried out with spectra of iron–argon comparison lamps.
In order to minimize the effects of atmospheric dispersion (Filippenko 1982), the spectrum was obtained as close to the meridian as
possible.

The second spectrum was obtained with the Kast double spectrograph (Miller \& Stone 1993) mounted on the 3\,m Shane telescope. 
Grism 600/4310 was used on the blue side and grating 300/7500 on the
red side, providing respective
resolutions of $\sim 5$\,\AA\ and $\sim 12$\,\AA\ and a range of 3600--10,700\,\AA.
The long 2\arcsec\ slit was aligned near the
parallactic angle (Filippenko 1982) to minimize
differential light losses caused by atmospheric dispersion,
and the exposure time was 1500\,s.

Flux calibration of both spectra was carried out with observations of Kitt Peak National Observatory standard
stars (Massey et al. 1988).

\subsubsection{Data Reduction}

One-dimensional (1D) spectra were extracted from the raw images by using the IRAF\footnote{IRAF is distributed by NOAO, which is operated by AURA, Inc., under cooperative agreement with the U.S. National Science Foundation (NSF).} package and standard 
procedures, including bias subtraction and flat-field correction. 
Both extracted 1D spectra were then calibrated in wavelength and in
flux with the corresponding comparison lamp and standard stars.
% \footnote{For the Kast spectrum, see Silverman et al.
% (2012) and Shivvers et al. (2019) for the details.}. 
The accuracy of the wavelength calibration is better than 1\,\AA\ for the Kast spectrum and better than 2\,\AA\ for the BFOSC spectrum. 
The telluric A-band (7600--7630\,\AA) and
B-band (around 6860\,\AA) absorption produced by atmospheric $\mathrm{O_2}$ molecules were removed from both spectra by using the
observation of the corresponding standard star.

Each calibrated  spectrum was then corrected for the Galactic extinction of 
$A_V=0.226$\,mag (Schlafly \& Finkbeiner 2011) taken from the NED. The correction was applied by
assuming the $R_V=3.1$ extinction law of our Galaxy (Cardelli et al. 1989).
Both spectra were then transformed to the rest frame.
The two resulting spectra are displayed in Figure 1 (lower two curves). 
% along with the correction of the relativity effect on the flux,
% according to the redshift.

\begin{figure}
   \centering
   \includegraphics[width=12cm]{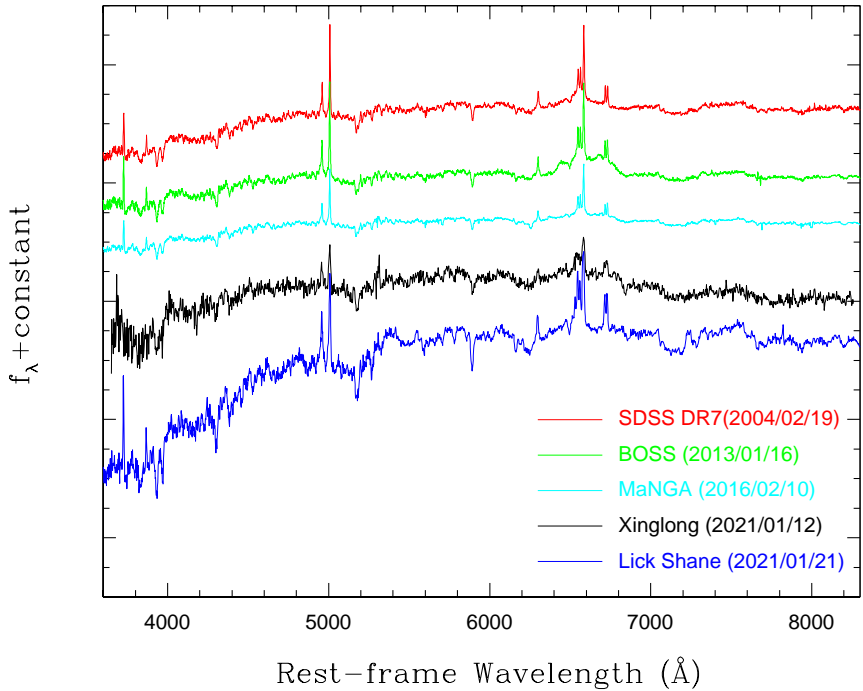}
   \caption{A comparison between the rest-frame spectra of B3\,0749+460A taken at different epochs. The two new spectra obtained with 
the Xinglong 2.16\,m telescope and the Shane 3\,m telescope are displayed by the lower two curves. The three spectra extracted from the 
SDSS DR7, BOSS, and MaNGA spectral database are shown by the upper three curves. The spectra are 
vertically shifted by an arbitrary amount for clarity.}
\label{Fig1}%
\end{figure}

\subsection{X-ray Follow-up Observation}

We proposed X-ray and ultraviolet (UV) follow-up observations of the object at the beginning of 2021 by using the Neil Gehrels {\it Swift} Observatory (Gehrels et al. 2004) X-ray telescope (XRT) and Ultraviolet/Optical Telescope (UVOT). 
The object was targeted (ObsID=00037357002) by XRT and UVOT simultaneously on 2021 February  12. The 
exposure times were 1400~s in  XRT Photon Counting (PC) mode and  1400~s for the UVOT. The UVOT image is, however, useless because of heavy contamination caused by the bright star in the field of view. 

The XRT data taken in the PC mode were reduced with HEASOFT version 6.27.2, along with the corresponding 
CALDB version 20190910.
The source spectrum was then extracted from the image in a circular region with a radius of 20.0\arcsec.
An adjacent region free of any sources was adopted to extract the background-sky spectrum. 
The corresponding ancillary response file was produced by the task \it xrtmkarf. \rm
The total XRT count rate in the 0.3--10\,keV  range was estimated to be $(4.62\pm0.44)\times10^{-2}\ \mathrm{count\ s^{-1}}$.

\section{Analysis and Results}

In order to identify the CL phenomenon in B3\,0749+460A and to reveal the underlying physics, 
spectral analysis was performed on the optical and X-ray spectra described in Section 2.

\subsection{Optical Spectroscopy: A CL-AGN}

Figure 1 compares the new spectra taken in 2021 with the previous spectra, which were obtained in different epochs, 
\rm extracted from the SDSS DR7, BOSS, and 
Mapping Nearby Galaxies at Apache Point Observatory (MaNGA) survey. 
The MaNGA spectrum is extracted from the datacube by an aperture with a diameter of 3\arcsec. 
At first glance, one can see clearly a significant variation of the H$\alpha$ line profile.  
The SDSS spectrum taken in 2013 shows the existence of a very broad, double-peaked H$\alpha$ 
component that is, however, absent or weak in the other four spectra. 
%A similar plot can be found in Figure 5 in Hon et al. (2020), where the 
%spectroscopy taken in 2016 by the Mapping Nearby Galaxies at Apache Point Observatory (MaNGA) survey is used instead.  
The following spectral analysis allows us to identify the object as a local repeat CL-AGN. 

\subsubsection{Subtraction of the Starlight Component}

We first remove the starlight component from each of the five spectra by 
modeling the stellar features with a linear combination of the first seven eigenspectra
through $\chi^2$ minimization, where the eigenspectra are built from the
standard single stellar population spectral library developed by Bruzual \& Charlot (2003). 
The minimization is carried out over the rest-frame wavelength range from 3700 to 8000~\AA, except for
the regions with strong emission lines --- Balmer lines (both
narrow and broad components), [O~III] $\lambda\lambda$4959, 5007,
[N~II] $\lambda\lambda$6548, 6583, [S~II] $\lambda\lambda$6716, 6731, [O~II] $\lambda$3727, 
[O~III] $\lambda$4363, and [O~I] $\lambda$6300.
Intrinsic extinction due to the host galaxy described by the Galactic extinction curve with $R_V=3.1$ is
also involved in the modeling. A potential featureless continuum contributed by the central AGN is ignored in 
the minimization; it is almost degenerate with the spectrum of early-type stars,
we do not focus on the stellar population in this study, and the effect caused by the featureless continuum is usually slight (e.g., Wang 2015).
The starlight templates are convolved with a fixed Gaussian profile in advance before the minimization for the 
Xinglong spectrum, because the observed absorption features are dominated by the instrumental profile. 
The subtraction of the starlight component is illustrated in Figure 2.

\begin{figure}
   \centering
   \includegraphics[width=12cm]{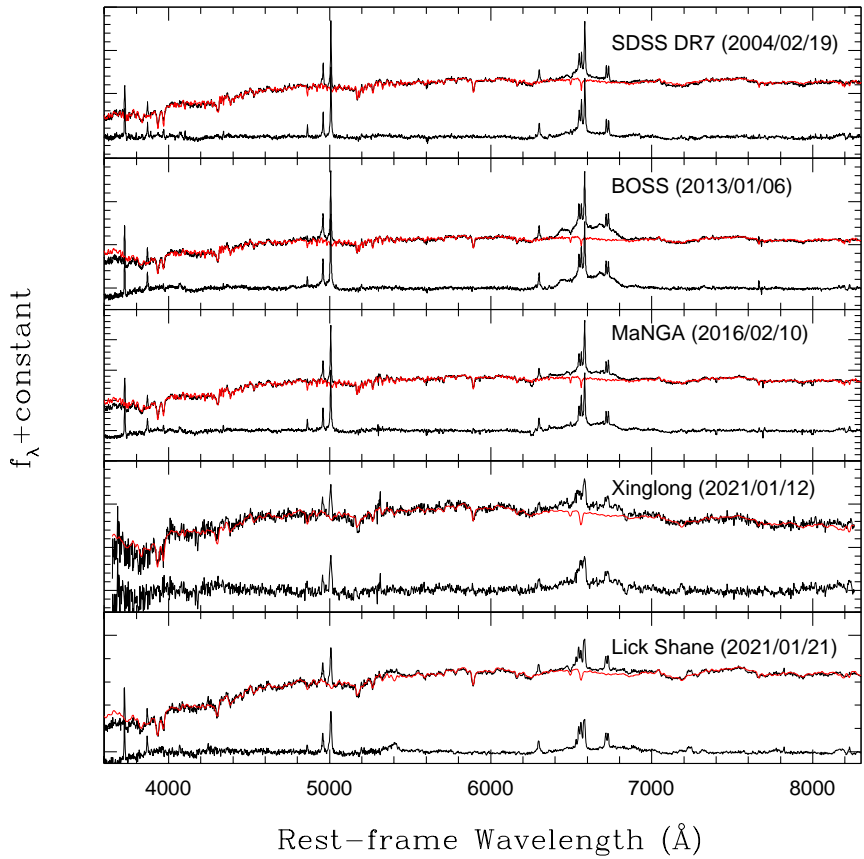}
   \caption{An illustration of the modeling and removal of the stellar continuum for the
five spectra (see the main text for the details). In each panel, the top black curve shows
the observed rest-frame spectrum, overplotted by the best-fit continuum indicated
by the red curve. The black curve underneath shows the continuum-removed emission-line 
spectrum.}
\label{Fig2}%
\end{figure}

\subsubsection{Line-Profile Modeling}

After removing the starlight component, we model the emission-line profiles in each of the spectra by a linear combination of a set of Gaussian
profiles in the H$\alpha$ and H$\beta$ regions by the SPECFIT task (Kriss 1994) in IRAF.
%\footnote{IRAF is distributed by the National Optical Astronomy Observatories, which are operated by the Association of Universities for Research in Astronomy, Inc., under cooperative agreement with the U.S. National Science Foundation (NSF)}. 
The 2021 Xinglong spectrum is not included in the profile modeling 
simply because the nearly contemporaneous 2021 Shane spectrum has better spectral resolution and higher signal-to-noise ratio (S/N).
In the modeling, the line-flux ratios of the [O~III] $\lambda\lambda$4959, 5007 and the [N~II] 
$\lambda\lambda$6548, 6583 doublets are fixed to their theoretical values of 1:3.
A sum of a narrow component and a blueshifted broad component is required to reproduce the 
[O~III] $\lambda\lambda$4959, 5007 line profiles 
in all three spectra (e.g., Boroson 2005; Zhang et al. 2013; Harrison et al. 2014; Woo et al. 2017; Wang et al. 2011, 2018a). 

The Balmer-line models are described as follows. 
\begin{itemize}
\item {\bf 2004 SDSS spectrum.} The H$\alpha$ line profile can be well reproduced by a combination of narrow (FWHM $\approx 400$ km~s$^{-1}$) and broad 
($\mathrm{FWHM} \approx 7800\ \mathrm{km\ s^{-1}}$) components\footnote{All line widths are 
not corrected for the instrumental spectral resolution throughout of the paper.}. A broad component is not needed to reproduce the profile of H$\beta$. 
\item {\bf 2013 BOSS spectrum.} In addition to a classical broad component with $\mathrm{FWHM} \approx 3000\ \mathrm{km\ s^{-1}}$, a double-peaked 
broad H$\alpha$ (e.g., Halpern, \& Filippenko 1988; Eracleous \& Halpern 1994; Strateva et al. 2003; Popovic et al. 2003) line containing 
redshifted and blueshifted broad components is necessary to properly reproduce the observed H$\alpha$ profile; thus, a total of three broad components is required.
Double-peaked Balmer emission lines are expected to 
originate from the rotating gas in an accretion disk (e.g., Chen et al. 1989; Popovic et al. 2004;
Bon et al. 2009a,b; Storchi-Bergmann et al. 2017). 

Initially, we model the H$\beta$ profile by a sum of a narrow and a broad H$\beta$ emission component;
however, this results in a residual on the blue side of [O~III]$\lambda4959$ line. Since the 
intrinsic profile of the [O~III]$\lambda4959$ line must be physically identical to that of the
[O~III]$\lambda5007$ line, the residual motivates 
us to reproduce the emission-line spectrum in the H$\beta$ region by an additional broad redshifted H$\beta$ emission component. 

We argue that both H$\alpha$ and H$\beta$ show the same double-peaked broad-line profiles in the 2013 SDSS spectrum, 
because of the comparable bulk velocities of the broad redshifted components. The bulk velocities 
are calculated to be $5980\pm90\,\mathrm{km\ s^{-1}}$ and $5310\pm140\,\mathrm{km\ s^{-1}}$ for H$\alpha$ and 
H$\beta$, respectively. The broad blueshifted H$\beta$ is likely undetectable owing to its weakness. 

\item {\bf 2016 MaNGA spectrum.} Except for its significant weakness,  the modeling of the H$\alpha$ line profile is similar to that of the 2013 BOSS spectrum.
The H$\beta$ emission line can, however, be modeled well by a single narrow component.  
 
\item {\bf 2021 Shane spectrum.} The classical H$\alpha$ broad component with a width of $\sim10^{3}\ \mathrm{km\ s^{-1}}$ is not necessary for 
reproducing the line profile.
The H$\alpha$ broad emission line can be described by a fainter double-peaked profile, compared to the MaNGA spectrum. 
%Bon et al. (2005, and reference therein) argued that the extended red component could result from a 
%gravitational redshift. 
Again, a single narrow Gaussian can reproduce the H$\beta$ line profile properly.
\end{itemize} 

The line-profile models are displayed in Figure 3 for the H$\beta$ ({\it left}) and
H$\alpha$ ({\it right}) regions, and the results are listed 
in Table 1.
All of the reported errors correspond to the 1$\sigma$
significance level and include only the uncertainties caused by the fitting,
not the removal of the stellar continuum.

To summarize,
in addition to a significant variation of the broad H$\alpha$ line profile, the spectral modeling and analysis enable us to reveal 
weak, broad (and most likely double-peaked) H$\beta$ components in the SDSS 2013 spectrum, 
but not in the other three spectra, suggesting that the object is a CL-AGN with 
type transitions of Sy1.9$\rightarrow$Sy1.8$\rightarrow$Sy1.9 
(Type 1.8 and 1.9 AGNs are classified according the existence and absence of a weak broad H$\beta$ line, respectively;
e.g., Osterbrock \& Ferland 2006.), although the prototypical CL refers to a transition between
Type 1 and Type 2 spectra.

\begin{figure}
   \centering
   \includegraphics[width=12cm]{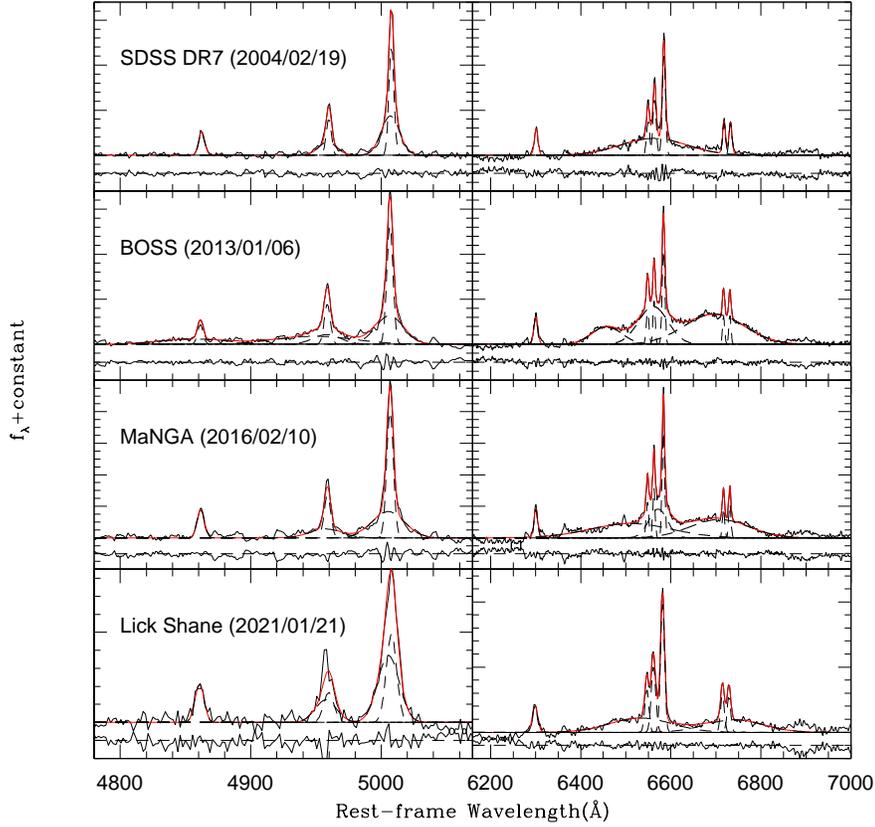}
   \caption{An illustration of the line-profile modeling with a linear combination of a set of Gaussian functions for H$\beta$ ({\it left panels}) and H$\alpha$ ({\it right panels}). 
In each panel, the modeled stellar continuum has already been removed from the original observed spectrum. The observed and modeled line profiles are plotted with the black and red
solid lines, respectively. Each Gaussian function is shown by a dashed line. The subpanel underneath the line spectrum presents the residuals between the observed
and modeled profiles. }
\label{Fig2}%
\end{figure}

\begin{table}
\renewcommand{\thetable}{\arabic{table}}
\centering
\caption{Results of Line-Profile Modeling and Analysis\label{chartable}}
\label{tab:decimal}
\footnotesize
\begin{tabular}{lcccc}
%\tablewidth{0pt}
\hline\hline
Parameters &  2004 Feb. 19 & 2013 Jan. 06 & 2016 Feb. 10 & 2021 Jan. 21 \\
(1) & (2) & (3) & (4) & (5)\\
\hline
\multicolumn{5}{c}{Line flux ($\mathrm{10^{-15}\ erg\ s^{-1}\ cm^{-2}}$)}\\
\hline
$F(\mathrm{[O~III]} \lambda5007)$ & $13.4\pm1.0$  & $16.3\pm0.4$ & $10.3\pm0.3$ & $25.1\pm4.1$ \\
$F(\mathrm{H\beta_b})$ & \dotfill & $3.7\pm0.4$ & \dotfill  &  \dotfill \\ 
$F(\mathrm{H\beta_{b2}})$ & \dotfill & $6.6\pm0.8$  & \dotfill &  \dotfill\\ 
$F(\mathrm{H\alpha_{b}})$ & $34.9\pm1.3$ & $34.2\pm0.6$ & $9.4\pm0.5$ & \dotfill \\
$F(\mathrm{H\alpha_{b1}})$ & \dotfill & $16.4\pm0.4$ & $20.4\pm0.5$ & $37.8\pm1.3$ \\
$F(\mathrm{H\alpha_{b2}})$ & \dotfill & $55.8\pm0.5$ & $18.9\pm0.3$ & $30.7\pm0.9$ \\ 
\hline
\multicolumn{5}{c}{Line width ($\mathrm{km\ s^{-1}}$)}\\
\hline
$\mathrm{FWHM(H\beta_b)}$ & \dotfill & $3760\pm400$ & \dotfill & \dotfill \\
$\mathrm{FWHM(H\beta_{b2})}$ & \dotfill &  $4260\pm520$ & \dotfill & \dotfill\\
$\mathrm{FWHM(H\alpha_{b})}$ & $7860 \pm250$ & $3500\pm60$ & $2210\pm80$ & \dotfill\\
$\mathrm{FWHM(H\alpha_{b1})}$ & \dotfill & $3640\pm80$ & $9760\pm220$ &  $7440\pm260$ \\
$\mathrm{FWHM(H\alpha_{b2})}$ & \dotfill & $7000\pm70$ & $6820\pm140$ &  $6660\pm30$ \\
\hline
$L/L_{\mathrm{Edd}}$ & 0.005 & 0.012 & 0.009 & 0.005\\
\hline
\end{tabular}
%\tablenotetext{a}{Only a total EW of CIV$\lambda\lambda$1548, 1550 doublets can be obtained.}
%\note{References: 1: XMM-SSC (2017); 2: Evans et al. (2020}
\end{table}

\subsubsection{Estimation of Black Hole Mass and Eddington Ratio}

After the line-profile modeling, we estimate the black hole mass ($M_{\mathrm{BH}}$) and Eddington ratio 
($L/L_{\mathrm{Edd}}$) in terms of the modeled H$\alpha$ broad emission line through the traditional method described by 
Wang et al. (2020a). Briefly, $M_{\mathrm{BH}}$ can be estimated by the calibration (Greene \& Ho 2007)
\begin{equation}
M_{\mathrm{BH}}=3.0\times10^6\bigg(\frac{L_{\mathrm{H\alpha}}}{10^{42}\ \mathrm{erg\ s^{-1}}}\bigg)^{0.45}\bigg(\frac{\mathrm{FWHM_{H\alpha}}}{1000\ \mathrm{km\ s^{-1}}}\bigg)^2\ M_\odot
\end{equation}
and $L/L_{\mathrm{Edd}}$ through a bolometric correction of $L_{\mathrm{bol}}=9\lambda L_{\lambda}(5100~{\rm \AA})$ (e.g., Kaspi et al. 2000), where (Greene \& Ho 2005)
\begin{equation}
\lambda L_\lambda(5100\AA)=2.4\times10^{43}\bigg(\frac{L_{\mathrm{H\alpha}}}{10^{42}\ \mathrm{erg\ s^{-1}}}\bigg)^{0.86}\ \mathrm{erg\ s^{-1}}.
\end{equation}

The 2004 SDSS spectrum is used as a reference in the subsequent comparison. 
In deriving the broad H$\alpha$ luminosity $L_{\mathrm{H\alpha}}$, the measured broad H$\alpha$
line fluxes in the BOSS, MaNGA, and Shane spectra are first scaled by
a factor determined by equaling the total [O~III] $\lambda$5007 line fluxes
to that of the reference one (e.g., Peterson et al. 2000). Then, 
the intrinsic extinction is corrected from the narrow-line flux ratio $\mathrm{H\alpha/H\beta}$ by 
assuming the Balmer decrement of standard Case B recombination and a
Galactic extinction curve with $R_V=3.1$. The value of $M_{\mathrm{BH}}$ is
estimated to be $\sim1.3\times10^8\,M_\odot$ based on the 2004 SDSS spectrum, which we use as
a fiducal value 
because of the regular and symmetric H$\alpha$ emission-line profile. With this value of $M_{\mathrm{BH}}$, 
the lowest row in Table 1 reports the estimated $L/L_{\mathrm{Edd}}$, where $L_{\mathrm{bol}}$ 
is estimated from the total H$\alpha$ broad emission. 
One can see a relation between $L/L_{\mathrm{Edd}}$ and spectral types.
%However, we argue that $L/L_{\mathrm{Edd}}$ is probably overestimated significantly for the 2021 Shane spectrum, since the H$\alpha$ line luminosity 
%is mainly contributed by the very wide red shoulder of the H$\alpha$ emission.   

\subsection{X-ray Energy Spectrum}
%\subsubsection{A Soft X-ray Spectrum in 2021}
\subsubsection{{\it Swift}/XRT  Spectra in 2008, 2016, and 2021} 

\begin{itemize}
\item {\it Swift/XRT 2021 spectrum.} Because of its low count rate, we attempt to model the X-ray energy spectrum of B3\,0749+460A by XSPEC (v12.11, Arnaud 1996) 
with two simple models over the 0.3--10~keV range in terms of the 
C-statistic (Cash 1979; Humphrey et al. 2009; Kaastra 2017). The adopted models can be  expressed as $wabs*zwabs*powerlaw$ 
and $wabs*zpcfabs*powerlaw$. In both cases, the Galactic hydrogen column density 
is fixed to be $N_{\mathrm{H}}=5.55\times10^{20}\ \mathrm{cm^{-2}}$ (Kalberla et al. 2005). 
%The best fit and its parameters are shown in Figure 4 (\bf the lower panel\rm) and Table 2, respectively. 
%All of the quoted uncertainties correspond to a 90\% significance level.
Both models return similar results characterized by a quite soft X-ray spectrum with a photon index of $\Gamma=2.8\pm0.5$. 

\item {\it Swift/XRT 2008 spectrum.} The object was observed by the {\it Swift}/XRT in pointing mode with an exposure time of 
9000\,s on 2008 February 24, and 
is included in the second {\it Swift}/XRT Point Source (2SXP) Catalog (1SXPS\,J075244.4+455655; Evans et al. 2020). 
The energy spectrum is extracted by the method described in Section 2.2, and analyzed by the same models applied to the 
2021 spectrum.  Both best-fitted models return a hard X-ray spectrum with a photon index of $\Gamma=1.7\pm0.1$. 
%Again, Figure 4 (the upper panel) and Table 2 show the best fit and the corresponding parameters.  

\item {\it Chandra 2016 spectrum.} The spectrum of the object was taken by ACIS-S onboard the {\it Chandra} satellite on 
2016 December 21. We reduced the data with software CIAO version 4.13 and the 
corresponding calibration database CALDB 4.9.5. The extracted spectrum is then again modeled by the XSPEC package,
which returns a quite hard power-law spectrum with a photon index of $\Gamma=1.5\pm0.1$.
In addition to the continuum models aforementioned, two Gaussian functions are required to model the weak 
Ar and Ca K$\alpha$ emission lines at 3.0 and 3.7\,keV, respectively.

\end{itemize} 

The best fits and their parameters are shown in Figure 4 and listed in Table 2.
In the table, all of the quoted uncertainties correspond to a 90\% significance level.

\begin{figure}
   \centering
\includegraphics[width=12cm]{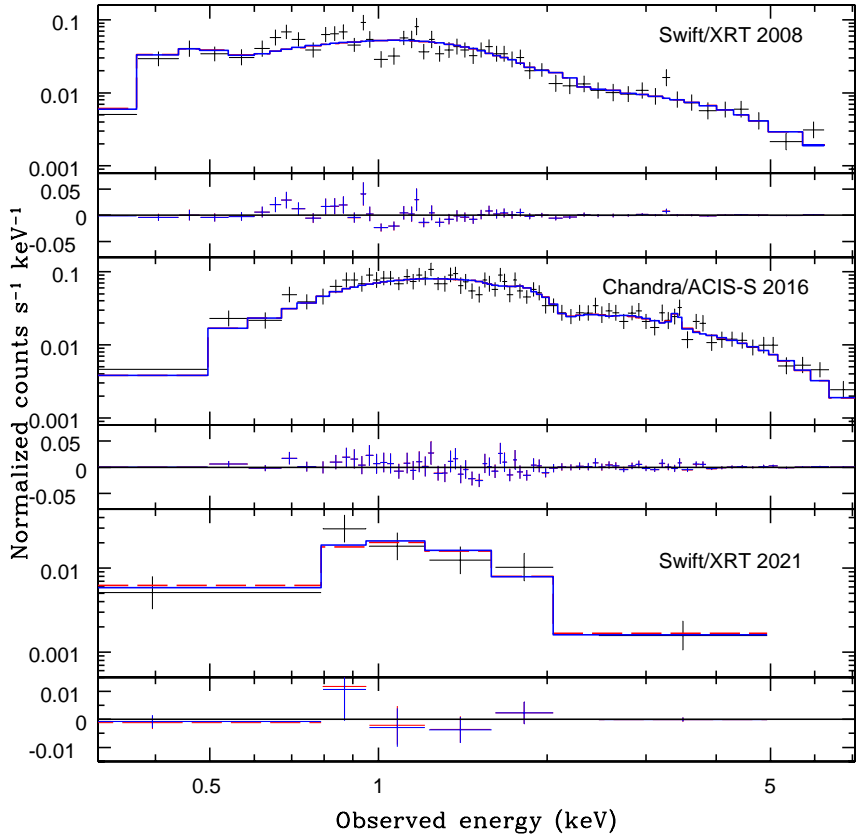}
\caption{X-ray spectrum of B3\,0749+460A taken at different epochs and the
best-fit spectral model expressed as $wabs*zwabs*powerlaw$ (Model 1, red-dashed line) and
$wabs*zpcfabs*powerlaw$ (Model 2, blue-solid line). The subpanel underneath the
spectrum shows the deviations, in units of $\mathrm{counts\ s^{-1}\ keV^{-1}}$, 
of the observed data from the best-fit model. 
}
\end{figure}

\begin{table}
\renewcommand{\thetable}{\arabic{table}}
\centering
\caption{X-ray spectra fit parameters of B3\,0749+460A.}
\label{tab:decimal}
\footnotesize
\begin{tabular}{llllll}
%\tablewidth{0pt}
\hline\hline
Parameter & \multicolumn{3}{c}{Value}  & Units & Description \\
    & \cline{1-3} & & & \\
     &  2008 & 2016  & 2021 & & \\
(1)  &   (2) & (3)  &(4) & (5)\\
\hline
\multicolumn{4}{l}{Model 1 - $wabs*zwabs*powerlaw$}\\
$N_{\mathrm{H}}$ & $0.01_{-0.01}^{+0.03}$ & $<0.01$ &  $0.25_{-0.09}^{+0.12}$ & $10^{22}\,\mathrm{cm^{-2}}$ & Local column density\\
$\Gamma$         & $1.65_{-0.11}^{+0.14}$ & $1.48_{-0.09}^{+0.10}$ & $2.75^{+0.54}_{-0.45}$  &  & Power-law index\\
$F$(2--10\,keV)  & $2.63_{-0.33}^{+0.36}\times10^{-12}$ & $1.50_{-0.18}^{+0.17}\times10^{-12}$  &   $3.70^{+0.28}_{-0.29}\times10^{-13}$ & $\mathrm{erg\ s^{-1}\ cm^{-2}}$ & Unabsorbed flux\\
Cash statistics  & 53.16/50 = 1.063 & 54.18/66 = 0.821 & 2.46/4 = 0.615 & & \\
\\
\multicolumn{4}{l}{Model 2 - $wabs*zpcfabs*powerlaw$}\\
$\eta_{\mathrm{H}}$ & $0.01_{-0.01}^{+1.33}$ &  $<0.01$ &  $0.30_{+0.64}^{-0.26}$  & $10^{22}\,\mathrm{cm^{-2}}$ & Local equivalent column density\\
$f$                 & 0.95 & \dotfill&  0.95  &  & Dimensionless covering fraction (fixed)\\
$\Gamma$            & $1.64_{-0.10}^{+0.15}$ & $1.48_{-0.09}^{+0.10}$ &  $2.76^{+1.01}_{-1.08}$  &  & Power-law index\\     
$F$(2--10\,keV)     & $2.65_{-0.32}^{+0.36}\times10^{-12}$ & $1.50_{-0.18}^{+0.17}\times10^{-12}$  & $3.84^{+0.27}_{-0.28}\times10^{-13}$ & $\mathrm{erg\ s^{-1}\ cm^{-2}}$ & Unabsorbed flux\\
Cash statistics     & 56.33/49 = 1.150 & 54.18/66 = 0.809 & 2.77/3 = 0.923 & & \\
\hline
\end{tabular}
%\tablenotetext{a}{Only a total EW of CIV$\lambda\lambda$1548, 1550 doublets can be obtained.}
%\note{References in the last column: (1) Ghirlanda et al. (2018); (2) Xue et al. (2019)}
\end{table}

\subsubsection{Spectral Slope Transition in the CL Phenomenon}

Here we compare the new X-ray observation with the previous ones to reveal the potential physical process occurring in the CL phenomenon of the object.   
%The object has been observed by the \it Swift/\rm XRT in 2008, and is included in the second Swift/XRT Point Source (2SXP) Catalog (1SXPS\,J075244.4+455655; Evans et al. 2020). 
 In addition to the 2SXP Catalog, the object is in the XMM-Newton Slew Survey Source Catalog (XMM-SSC, XMMSL2\,J075244.7+455654). 
There are three detections at different epochs in the XMM-SSC catalog from 2005 to 2011.

The comparison is shown in Table 3. Columns (2) and (3) list the intrinsic fluxes in the 0.3--10~keV ($F$(0.3-–10)\,keV) and the observed flux in the 
0.2--12~keV ($F$(0.2-–12)\,keV) bands, respectively.
In order to compare with the previous values, the tabulated fluxes in 2021 are obtained from our
fitting with the model of $wabs*zwabs*powerlaw$. The XMM-Newton hardness ratio\footnote{The 
XMM-Newton hardness ratio is defined as $\mathrm{(R_{b7}-R_{b6})/(R_{b7}+R_{b6})}$, where 
$R_{\mathrm{b7}}$ and $R_{\mathrm{b6}}$ are the hard (2--12~keV) and soft (0.2--2~keV) 
band rates, respectively.} 
is listed in Column (4).  
Columns (5) and (6) are the hardness ratios HR1 and HR2 defined by Evans et al. (2020) for the \it Swift \rm mission\footnote{The \it Swift \rm hardness ratios HR1 and HR2 are 
defined as $\mathrm{HR1}=(M-S)/(M+S)$ and $\mathrm{HR2}=(H-M)/(H+M)$, where $S$, $M$, and $H$ are the count rates in the 0.3--1~keV, 1--2~keV, and 2--10~keV energy bands, 
respectively.}.
All of the uncertainties shown in the table correspond to a 90\% significance level, after taking into account proper error
propagation. For the 2008 {\it Swift}/XRT observation, the hardness ratios obtained in the current work are highly consistent with the 
values reported in the 2SXP Catalog (Evans et al. 2020).

The upper panel in Figure 5 shows the long-term variation of the X-ray flux and hardness ratio $\mathrm{HR1_{XMM}}$ of 
the object. On the one hand, as shown in the table, the X-ray flux decreases gradually by more than an order of magnitude 
during the period from 2005 to 2021, which does not follow the spectral-type transition. On the other hand,
based on the hardness ratios at different epochs, 
one can see from the comparison that the CL phenomenon identified in B3\,0749+460A tends to be related to the X-ray slope.  
The X-ray spectrum is found to be soft ($\mathrm{HR1=-0.50\pm0.18}$ in 2005 and $\mathrm{HR1=-0.41\pm0.16}$ in 2021)  at the ``turn-off'' state with a Seyfert 1.9-like spectrum, and to be hard ($\mathrm{HR1=0.21\pm0.35}$) in 2011 when the object was possibly 
at  the ``turn-on'' state with a Seyfert 1.8-like  spectrum.

\begin{table}
\renewcommand{\thetable}{\arabic{table}}
\centering
\caption{Comparison of X-ray emission}
\label{tab:decimal}
\footnotesize
\begin{tabular}{ccccccccccc}
%\tablewidth{0pt}
\hline\hline
Year of Obs. & $F$(0.3--10)\,keV & $F$(0.2--12)\,keV & $\mathrm{HR1_{XMM}}$ & 
$\mathrm{HR1_{Swift}}$ &  $\mathrm{HR2_{Swift}}$ & Sp. type & Mission & Reference\\
         &          \multicolumn{2}{c}{($10^{-12}\ \mathrm{erg\ s^{-1}\ cm^{-2}})$}  & & &  & & & & \\ 
         \cline{2-3}
(1) & (2) & (3) & (4) & (5) & (6) & (7) & (8)& (9)\\
\hline
2005 & \dotfill & $10.3\pm1.9$ & $-0.50\pm0.18$ &  \dotfill & \dotfill & Sy1.9 & XMM-Newton & 1\\
2006$^a$ & \dotfill & $6.7\pm1.8$ & $\dotfill$ &  \dotfill & \dotfill & \dotfill & XMM-Newton & 1\\
2008 & $5.2\pm0.3$ & \dotfill  &  \dotfill              &  $0.07\pm0.04$    &   $-0.09\pm0.04$ & \dotfill & {\it Swift} & 2 \\
     & $4.3\pm0.3$ & $4.3\pm0.4$ & $-0.35\pm0.03$ & $0.07\pm0.04$ & $-0.08\pm0.04$  & \dotfill & {\it Swift} & This work \\
2011 & \dotfill & $6.7\pm2.3$   & $0.21\pm0.35$  &  \dotfill & \dotfill & Sy1.8 & XMM-Newton & 1\\
2016 & $2.2\pm0.2$ & $2.3\pm0.2$ & $-0.43\pm0.03$ & $-0.05\pm0.03$ & $-0.12\pm0.04$ & Sy1.9 & {\it Chandra} & This work \\
2021 & $2.1^{+1.0}_{-0.8}$  &  $0.8^{+0.6}_{-0.3}$ & $-0.41\pm0.16$ & $-0.42\pm0.14$ & $-0.27\pm0.20$ & Sy1.9 & {\it Swift} & This work\\
\hline
\end{tabular}
\tablenotetext{a}{There is no reported value of HR1 in the catalog due to the lack of the count rate in
the B7 channel.}
\note{References: 1: XMM-SSC (2017); 2: Evans et al. (2020)}
\end{table}

\begin{figure}
\centering
\includegraphics[width=12cm]{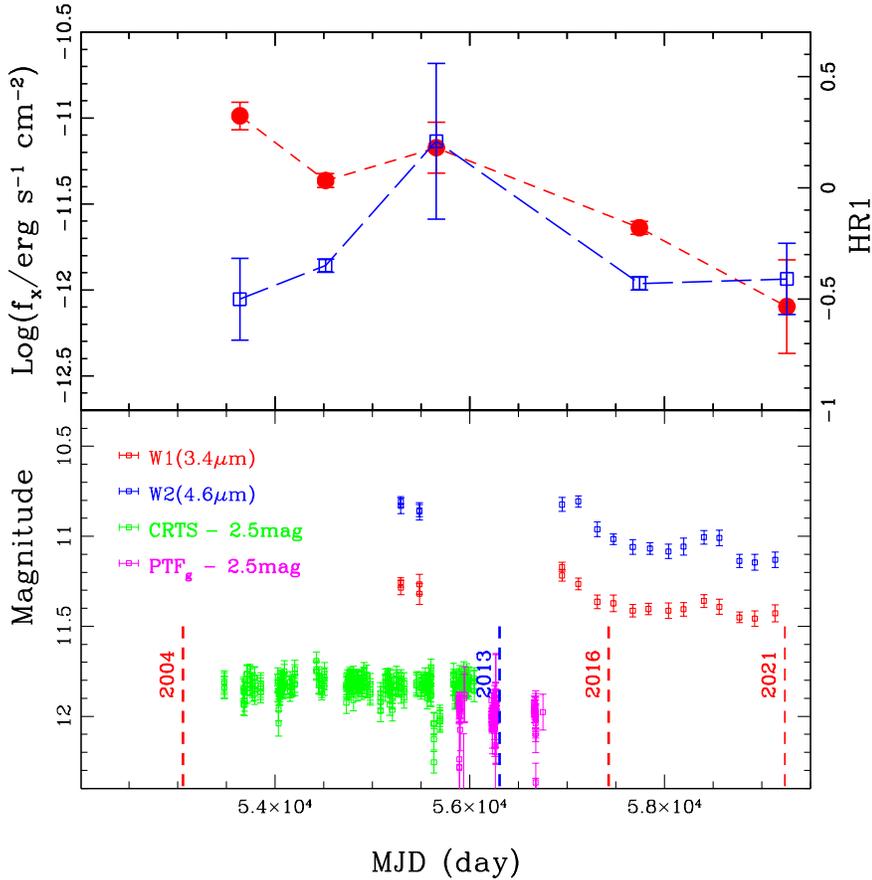}
\caption{\it Upper panel: Evolution of the observed X-ray flux in 0.2-12keV (red symbol) and 
XMM-Newton HR1 (blue symbol). \it Lower panel: \rm Multi wavelength light curves of B3\,0749+460A detected by the Catalina Real-time Transient Survey (Drake et al. 2011),
Palomar Transient Factory survey (Law et al. 2009)and {\it WISE} . \rm 
The light curves are binned by averaging the measurements within one day.
The vertical dashed lines mark the epochs of optical spectra, where the blue line denotes
the ``turn-on'' state with a Seyfert 1.8-like spectrum having weak broad H$\beta$ emission lines,
and the red ones the ``turn-off'' state with a Seyfert 1.9-like spectrum. The corresponding years are 
marked beside the vertical lines.}
\end{figure}

\section{Conclusions and Discussion}

We identify B3\,0749+460A as a new local CL-AGN by comparing SDSS spectra previously taken 
in 2004, 2013, and 2016 with new spectra taken by us in 2021. The object shows type transitions of Sy1.9$\rightarrow$Sy1.8$\rightarrow$Sy1.9. In particular, the classical broad H$\alpha$ component fades 
away from 2013 to 2021. 
A follow-up observation in X-rays taken by the {\it Swift}/XRT 
enables us to reveal (1) gradually decreasing X-ray emission since 2005, and (2) the X-ray spectrum 
is soft in the optical ``turn-off'' state with a Seyfert 1.9-like spectrum. Also, the X-ray spectrum 
is hard in the `turn-on'' state with a Seyfert 1.8-like spectrum, although there is a separation of about two years between the optical and X-ray 
observations.

\subsection{Broad H$\beta$ Emission}
Here we argue that the identified spectral-type transitions are not caused by the removal of the stellar continuum. 
The left panel in 
Figure 6 shows the differential spectra of the object obtained by adopting the 2004 SDSS spectrum as the reference, after scaling the others according to the total [O~III] $\lambda$5007 line flux.
One can see that, in addition to the redshifted H$\alpha$ broad component, 
there is a bump at the red side of H$\beta$ in the 2013 BOSS spectrum. Three arguments support our belief that the 
bump represents the redshifted H$\beta$ broad component.
First, as shown in the right panel of the figure, the two bumps exhibit quite similar profiles, after taking into 
account a measured bump flux ratio of 3 that is comparable to the average Balmer ratio of the broad Blamer lines in AGNs (Dong et al. 2006). 
Second, based on simple Gaussian fitting,
the two bumps show comparable 
bulk redshift velocities with respect to the corresponding rest-frame wavelength. The measured bulk velocities
are $\Delta v = 5895\pm87\,\mathrm{km\ s^{-1}}$ and $5310\pm140\,\mathrm{km\ s^{-1}}$ for H$\alpha$ and H$\beta$, respectively. Finally, the bump in the H$\beta$ region is unlikely to result from 
He~I $\lambda4922$ broad emission (e.g., Veron et al. 2002), simply because the 
broad He~I $\lambda$5876 emission line is not detected in the observed and differential spectra.

\begin{figure}
   \centering
\includegraphics[width=12cm]{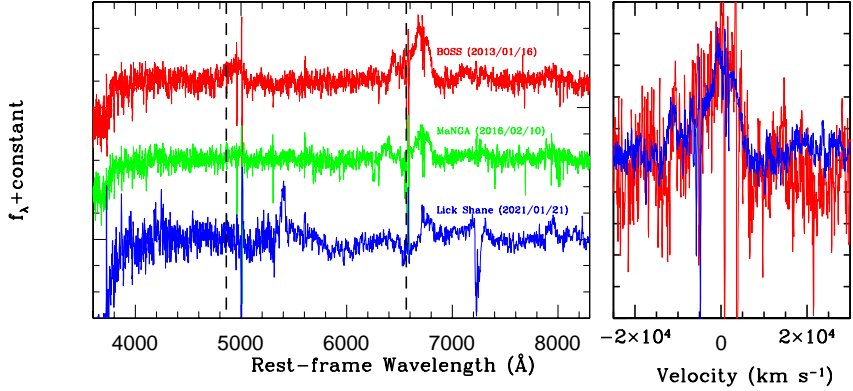}
\caption{
{\it Left panel:} A comparison of the differential spectra taken in 2013, 2016, and 2021, when the SDSS 2004 spectrum is used as a reference. Before the subtraction, the two spectra are scaled in 
flux level according to their total [O~III] $\lambda5007$ line fluxes. The two vertical 
lines mark the rest-frame wavelengths of the H$\beta$ and H$\alpha$ emission lines. One can 
see clearly a redshifted broad H$\beta$ component in the BOSS 2013 spectrum, but not in the 
2021 Shane spectrum. In the Shane spectrum, the features around 5400\,\AA\ are artificial,
caused by uncertainty in the low-S/N region where the blue-side and red-side CCD spectra are combined.
{\it Right panel:} A comparison of profiles of the two redshifted bumps, after scaling them by
a measured flux ratio of the two bumps of 3 (red, H$\beta$; blue, H$\alpha$).
}
\end{figure}

\subsection{A Comparison with Hon et al.}
Although our new spectrum taken in 2021 is comparable to the 2016 MaNGA spectrum shown by Hon et al. (2020),
our current study is quite improved in the spectral analysis when compared to Hon et al. (2020). The stellar-component subtraction, 
line-profile modeling, and a comparison within the differential spectrum enable us to argue for the existence of a broad H$\beta$ line in the 
2013 SDSS spectrum, which suggests that a CL phenomenon occurred in the object. Hon et al. (2020) argued against a typical
CL phenomenon in the object, because of the short variation timescale of $\sim1000$\,d. 
However, a CL timescale of $\sim1$--3\,yr has been revealed in previous investigations (e.g., Merloni et al. 2015; Runnoe et al. 2016; Macleod et al. 2016).  
In addition,  the supernova (SN) explanation of the SDSS 2013 spectrum suggested by Hon et al. (2020) can be 
potentially excluded by the lack of Ca~II triplet emission in the spectrum.   
Finally, taking into account the 2016 MaNGA spectrum,
the object experienced type transitions of Sy1.9$\rightarrow$Sy1.8$\rightarrow$Sy1.9 within a duration of 13\,yr (from 2004 to 2016), 
although the broad H$\alpha$ emission is found to be further weaker in 2021 than in 2016.

\subsection{Physical Implications}

The lower panel in Figure 5 shows the multiwavelength light curves of B4\,0749+460A. 
The mid-infrared (MIR) light curves in
the $w1$ and $w2$ bands detected by the {\it Wide-field Infrared Survey Explorer} 
({\it WISE} and {\it NEOWISE-R}; Wright et al. 2010; Mainzer et al. 2014) indicate that the 
brightness in both bands decreases gradually by $\sim0.3$\,mag when the object
changes from a Seyfert 1.8 to a Seyfert 1.9 nucleus, consistent with the expectation
of the accretion-rate enhancement scenario of the CL phenomenon 
(e.g., Sheng et al. 2017; Stern et al. 2018; Wang et al. 2019).
The possibility of a normal SN mentioned above can be further decreased according to the observed MIR variability with an
absolute magnitude in the $w2$ band of $-23.6$\,mag. This value is marginally larger than the brightest one (i.e., $-22.8$\,mag at 4.5\,$\mu$m 
for SN\,2010jl) reported by Szalai et al. (2019), who studied a sample of MIR light curves of hundreds of SNe.

With increasing cases of CL-AGNs, a few possible models have recently been proposed (e.g., Wang \& Bon 2021).
Based on the statistical studies performed by MacLeod et al. (2019) and Wang et al. (2019, 2020a,b),
there is some evidence 
for the CL phenomenon being understood by the disk-wind BLR model.
In the model,  a classical BLR can be sustained if $L/L_{\mathrm{Edd}}$ is above 
a critical value of $\sim 10^{-6}$--$10^{-3}$, when the fiducal values of a set of
parameters of the disk are adopted (e.g., Nicastro 2000; Elitzur \& Ho 2009). 
In B3\,0749+460A, when the $L/L_{\mathrm{Edd}}$ listed in Table 1 decreases gradually from 0.012 to 0.005,
the classical broad H$\alpha$ component in fact fades away from 2013 to 2021.
This link between the strength of the classical H$\alpha$ broad component and $L/L_{\mathrm{Edd}}$ can be reinforced if we instead estimate $L_{\mathrm{bol}}$ from the X-ray emission.
Our X-ray analysis suggests a decreasing $L/L_{\mathrm{Edd}}$ from 0.017 in 2011 to 0.002 in 2021,
when a bolometric correction of $L_{\mathrm{bol}}=16L_{\mathrm{X}}$ is adopted in the calculation. 
In addition, it is well known that the double-peaked broad Blamer lines observed frequently in AGNs can originate from gas in a single accretion disk (e.g., Storchi-Bergmann et al. 2017, and references therein).   
The double-peaked broad H$\alpha$ component is found to weaken synchronously with the classical broad H$\alpha$ in the object, implying a disk origin for the classical broad H$\alpha$
component. Alternatively, 
Pan et al. (2021) and J. Feng et al. (2021) show that the observed repeat CL-AGNs could be explained by reducing the 
disk burst period by including the effect of a large-scale magnetic field (e.g., Dexter \& Begelman 2009)
in the disk-instability model suggested by Sniegowska et al. (2020).
%A quite low $L/L_{\mathrm{Edd}}$ value is actually estimated 
%for B3\,0749+460A, especially in the ``turn-off'' state. 

%We therefore instead estimate the 2021 bolometric luminosity from the X-ray emission. 
%Based on the 2021 \it Swift/\rm XRT observation, 
%the unabsorbed X-ray luminosity in the 2--10~keV band is $L_{\mathrm{X,2-10\,keV}}=2.5\times10^{42}\ %\mathrm{erg\ s^{-1}}$, which yields a $L/L_{\mathrm{Edd}}\approx0.003$ when a bolometric
%correction of $L_{\mathrm{bol}}=16L_{\mathrm{X}}$ is adopted in the calculation.  

A transition of the accretion state, similar to that seen in XRBs, has been proposed recently as an explanation 
of the CL phenomenon (e.g., Ruan et al. 2019; Ai et al. 2020), although it seems that this scenario might not be generally applicable. 
%The relationship between accretion state and X-ray spectral slope, which has been 
%well established in XRBs, is in fact contrary to that identified here in B3\,0749+460A. 
The CL phenomenon in 
UGC\,3223, a local repeat CL-AGN, is found to be unrelated 
to the X-ray hardness ratio, but instead related to the X-ray luminosity (Wang et al. 2020b). 
Following Ai et al. (2020), the hardness ratio $\mathrm{CR_{soft}/(CR_{soft}+CR_{hard})}$ is plotted against $L/L_{\mathrm{Edd}}$ 
in Figure 7 for B3\,0749+460A, where $L/L_{\mathrm{Edd}}$ is derived from the unabsorbed X-ray 2--10\,keV luminosity by assuming a bolometric correction of $L_{\mathrm{bol}}=16L_{\mathrm{X}}$. After converting 
the photon index $\Gamma$ to the hardness ratio, the $\Gamma-L/L_{\mathrm{Edd}}$ relationships for low-luminosity AGNs (Constantin et al. 2009)
and bright AGNs (Risaliti et al. 2009) are overplotted in the figure, which stand for an 
advection-dominated accretion flow (ADAF) at low 
$L/L_{\mathrm{Edd}}$ and a standard accretion disk at high $L/L_{\mathrm{Edd}}$, respectively.
The $\Gamma-L/L_{\mathrm{Edd}}$ relationships can be well understood by the change of the 
Compton $y$ parameter with the released energy (e.g., Esin et al. 1997; Janiuk \& Czerny 2000). 
One can see from the figure that
the evolution of the object likely follows the ADAF locus and does not follow the V-shape transitions revealed by Ai et al. (2020). In fact, assuming the 
aforementioned disk-wind model, the (dis)appearance of a classical BLR is sensitive to accretion (wind) luminosity in individual AGNs, suggesting that a change of accretion mode is possibly a necessary 
(rather than a sufficient) condition for the CL phenomenon. 

\begin{figure}
   \centering
\includegraphics[width=12cm]{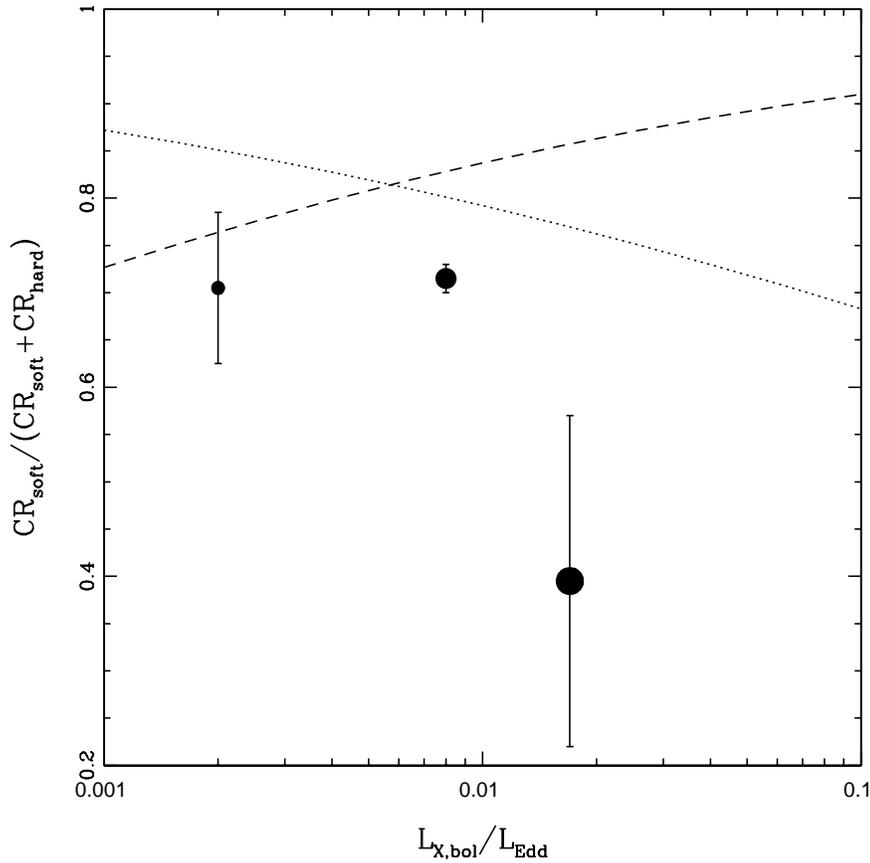}
\caption{Hardness ratio plotted against the $L/L_{\mathrm{Edd}}$ for B3\,0749+460A. 
$\mathrm{CR_{soft}}$ and $\mathrm{CR_{hard}}$ denote the count rates at 0.2--2 and 2--12\,keV, respectively.
$L/L_{\mathrm{Edd}}$ is estimated from the unabsorbed 2--10\,keV luminosity after assuming a bolometric
  correction of $L_{\mathrm{bol}} = 16L_{\mathrm{X}}$. The symbol size is proportional to the
  broad H$\alpha$ emission. The $\Gamma-L/L_{\mathrm{Edd}}$ relationships of low-luminosity AGNs and 
bright AGNs are denoted by the dotted and dashed lines, respectively. 
\rm
}
\end{figure}

%With increasing cases of repeat CL-AGNs, a few possible models have been proposed in past year.  
%On the one hand, the theoretical study by Wang \& Bon (2020) suggests that CL-AGNs could be triggered by 
%close binaries of SMBHs with a high eccentricity. A tidal torque on the mini-disk of each SMBH can either 
%squeeze or expand the disk, resulting in a spectral type transition cycle determined by the orbital period.    
%The authors argue that this scenario is able to explain the highly asymmetric and double-peaked broad-line
%profiles observed in CL-AGNs, such as the current object B3\,0749+460A whose 
%CL phenomenon is in fact found to tightly relate to the significant variations of the double-peaked Balmer-line components. 

%A larger spectroscopy sample is therefore needed to address this issue, and that is the strength of \it SVOM \rm, thanks to its capability of rapid identification of optical candidates of afterglow and its anti-solar pointing strategy.

\begin{acknowledgements}
The authors thank the anonymous referee for a careful review and helpful suggestions that greatly improved the manuscript.
This study is supported by the National Natural Science Foundation of China under grant 11773036, and by the Strategic Pioneer Program on Space Science, Chinese Academy of Sciences, grants XDA15052600 and XDA15016500.
The authors are grateful for support from the National Key Research and Development Project of China (grant 2020YFE0202100). 
J.W. is supported by the Natural Science Foundation of Guangxi (2020GXNSFDA238018) and by the Bagui Young Scholars Program. 
This study is supported by the Open Project Program of the Key Laboratory of Optical Astronomy, NAOC, CAS.
A.V.F.'s group is supported by the Christopher R. Redlich Fund and the Miller
Institute for Basic Research in Science (in which A.V.F. is a Senior Miller Fellow).
We thank the {\it Swift} Acting PI, Brad Cenko, for approving our target-of-opportunity request, as well as the {\it Swift} observation team for assistance.  This study uses the NASA/IPAC Extragalactic
Database (NED), which is operated by the Jet Propulsion
Laboratory, California Institute of Technology.
It also uses data collected by the {\it Wide-field Infrared Survey Explorer (WISE)},
which is a joint project of the University of California, Los
Angeles, and the Jet Propulsion Laboratory/California Institute
of Technology, funded by the National Aeronautics and Space
Administration (NASA).

%JW \& DWX are supported by the National Natural Science Foundation of China under grants
%11773036 and 11473036.
%MZK is supported by NSFC Youth Foundation (No. 11303008) and by Astronomical Union Foundation under grant
%No. U1831126.
%This study is supported by the National Basic
%Research Program of China (grant 2014CB845800), the NSFC under grants 11533003, and the Strategic
%Pioneer Program on Space Science, Chinese Academy of Sciences, Grant No.XDA15052600.
%The study is supported by the National Basic Research Program of China
%(grant 2009CB824800).  
%This study uses the SDSS archive data that was created and distributed by the Alfred P.
%Sloan Foundation, the Participating Institutions, the National Science
%Foundation, and the U.S. Department of Energy Office of Science.

\end{acknowledgements}

\label{lastpage}


\begin{thebibliography}{}
\bibitem[Ai et al. (2020)]{Ai20} Ai, Y. L., Dou, L. M., Yang, C. W., et al. 2020, \apjl, 890, 29 
\bibitem[Antonucci (1993)]{ant93} Antonucci, R. R. J. 1993, \araa, 31, 473
\bibitem[Arnaud (1996)]{arn96} Arnaud, K. A. 1996, ASPC, 101, 17 
\bibitem[Bon et al. (2009a)]{bon09a} Bon, E., Gavrilovic, N., La Mura, G., \& Popovic, L. C. 2009a, NewAR, 53, 121  
\bibitem[Bon et al. (2005)]{bon05} Bon, N., Bon, E., Marziani, P., \& Jovanovic, P. 2015, \apss,360, 41 
\bibitem[Bon et al. (2009b)]{bon09b} Bon, E., Popovic, L. C., Gavrilovic, N., La Mura, G., \& Mediavilla, E. 2009b, \mnras, 400, 924
\bibitem[Boroson (2005)]{bor05} Boroson, T. A. 2005, \aj, 130, 381
\bibitem[Bruzual \& Charlot (2003)]{brc03} Bruzual, G., \& Charlot, S. 2003, \mnras, 344, 1000
\bibitem[Cardelli et al. (1989)]{car89} Cardelli, J. A., Clayton, G. C., \& Mathis, J. S. 1989, \apj, 345, 245
\bibitem[Cash (1979)]{cas79} Cash, W. 1979, \apj, 228, 939
\bibitem[Chen et al. (1989)]{che89} Chen, K., Halpern, J. P., \& Filippenko, A. V. 1989, \apj, 339, 742 
\bibitem[Constantin et al. (2009)]{con09} Constantin, A., Green, P., Aldcroft, T., Kim, D. -W, Haggard, D, Barkhouse, W., \& Anderson, S. F. 2009, \apj, 705, 1336 
\bibitem[Dexter \& Begelamn (2019)]{deb19} Dexter, J., \&  Begelman, M. C. 2019, \mnras, 483, L17  
\bibitem[Drake et al. (2011)]{dra11} Drake, A. J., Djorgovski, S. G., Mahabal, A., et al.  2011,
  %The catalina real-time transient  survey.
  in New Horizons in Time-Domain Astronomy,
  ed. E. Griffin, R. J. Hanisch, \& R. L. Seaman (Proceedings of
  the International Astronomical Union, Vol. 7, No. S285),
  306. 
% \bibitem[Denney et al. (2014)] {den14} Denney, K. D., De Rosa, G., Croxall, K., et al. 2014, \apj, 796, 134
\bibitem[Elitzur \& Ho (2009)]{elh09} Elitzur, M., \& Ho, L. C. 2009, \apjl, 701, 91
\bibitem[Elitzur \& Shlosman (2006)]{els06} Elitzur, M., \& Shlosman, I. 2006,  \apjl, 648, 101
\bibitem[Eracleous \& Halpern (1994)]{era94} Eracleous, M., \& Halpern, J. P. 1994, \apjs, 90, 1
\bibitem[Esin et al. (1997)]{esi97} Esin, A. A., McClintock, J. E., \& Narayan, R. 1997, \apj, 489, 865 
\bibitem[Evans et al. (2020)]{eva20} Evans, P. A., Page, K. L., Osborne, J. P., et al. 2020, \apjs, 247, 54
\bibitem[Fan et al. (2016)]{fan16} Fan, Z., Wang, H., Jiang, X., et al. 2016, \pasp, 128, 115005
\bibitem[Feng et al. (2021)]{h-fen21} Feng, H.-C., Hu, C., Li, S.-S., et al. 2021, \apj, 909, 18
\bibitem[Feng et al. (2021)]{j-fen21} Feng, J. J., Cao, X. W., Li, J. W., \& Gu, W. M., 2021, \apj, 961, 61
\bibitem[Filippenko (1982)]{fil82} Filippenko, A. V. 1982, \pasp, 94, 715
\bibitem[Frederick et al. (2019)]{fre19} Frederick, S., Gezari, S., Graham, M. J., et al. 2019, \apj, 883, 31
\bibitem[Gehrels et al. (2004)]{geh04} Gehrels, N., Chincarini, G., Giommi, P., et al. 2004, \apj, 611, 1005  
\bibitem[Gezari et al. (2017)]{gez17} Gezari, S., Hung, T., Cenko, S. B., et al. 2017, \apj, 835, 144
\bibitem[Graham et al. (2020)]{gra20} Graham, M. J., Ross, N. P., Stern, D., et al. 2020, \mnras, 491, 4925
% \bibitem[Graham et al. (2019)]{gra19} Graham, M. J., Ross, N. P., Stern, D., et al. 2020, \mnras, 491, 4925
\bibitem[Greene \& Ho (2005)]{grh05} Greene, J. E., \& Ho, L. C. 2005, \apj, 630 ,122
\bibitem[Greene \& Ho (2007)]{grh07} Greene, J. E., \& Ho, L. C. 2007, \apj, 670, 92
\bibitem[Grimm et al. (2003)]{gri03} Grimm, H. -J., Gilfanov, M., \& Sunyaev, R. 2003, \mnras, 339, 793
\bibitem[Guo et al. (2019)]{guo19} Guo, H. X., Sun, M. Y., Liu, X., Wang, T. G., Kong, M. Z., Wang, S., Sheng, Z. F., \& He, Z. C. 2019, \apjl, 833, 44 
\bibitem[Halpern \& Filippenko (1988)]{hf88} Halpern, J. P., \& Filippenko, A. V. 1988, Nature, 331, 46
\bibitem[Hao et al. (2011)]{hao11} Hao, C.-N., Kennicutt, R. C., Johnson, B. D., Calzetti, D., Dale, D. A., \& Moustakas, J. 2011, \apj, 741, 124 
\bibitem[Harrison et al. (2014)]{har14} Harrison, C. M., Alexander, D. M., Mullaney, J. R., \& Swinbank, A. M. 2014, \mnras, 441, 3306
\bibitem[Hon et al. (2020)]{hon20}  Hon, W. J., Webster, R., \& Wolf, C. 2020, \mnras, 497, 192 
% \bibitem[Husemann et al. (2016)]{hus16} Husemann, B., Urrutia, T., Tremblay, G. R., et al. 2016, \aap, 593, L9
\bibitem[Huffman (1977)]{huf77} Huffman, D. R. 1977, AdPhy, 26, 129
\bibitem[Humphrey et al. (2009)]{hum09} Humphrey, P. J., Liu, W. H., \& Buote, D. A. 2009, \apj, 693, 822  
\bibitem[Hutsemekers et al. (2019)]{hut19} Hutsemekers D., Agis Gonzalez B., Marin F., Sluse D., Ramos Almeida C., Acosta Pulido J.-A., 2019, \aap, 625, A54
% \bibitem[Jiang et al. (2016)]{jia16} Jiang, Y. F., Davis, S. W., \& Stone, J. M. 2016, \apj, 827, 10
\bibitem[Janiuk \& Czerny (2000)]{jac00} Janiuk, A., \& Czerny, B. 2000, NewA, 5, 7  
\bibitem[Kaastra (2017)]{kaa17} Kaastra, J. S. 2017, \aap, 605, 51 
\bibitem[Kalberla et al. (2005)]{kal05} Kalberla, P. M. W., Burton, W. B., Hartmann, D., Arnal, E. M., Bajaja, E., Morras, R., \& Poppel, W. G. L. 2005, \aap, 440, 775 
\bibitem[Kaspi et al. (2000)]{kas00} Kaspi, S., Smith, P. S., Netzer, H., et al. 2000, \apj, 533, 631
\bibitem[Kollatschny et al. (2020)]{kol20} Kollatschny, W., Grupe, D., Parker, M. L., et al. 2020, \aap, 638, 91
\bibitem[Kollatschny et al. (2018)]{kol18} Kollatschny, W., Ochmann, M. W., Zetzl, M., Haas, M., Chelouche, D., Kaspi, S., Pozo Nuñez, F., \& Grupe, D. 2018, \aap, 619, 168 
\bibitem[Kriss (1994)]{kri94} Kriss, G. 1994, in ASP Conf. Ser. 61, Astronomical Data Analysis Software
and Systems III, ed. D. R. Crabtree, R. J. Hanisch, \& J. Barnes (San Fransisco, CA: ASP), 437
\bibitem[LaMassa et al. (2015)]{lam15} LaMassa, S. M., Cales, S., Moran, E. C., et al. 2015, \apj, 800, 144
\bibitem[Law et al. (2009)]{law09} Law, N. M., Kulkarni, S. R., Dekany, R. G., et al. 2009, \pasp, 121, 1395   
\bibitem[Lawrence (2018)]{law18} Lawrence, A. 2018, \nat\ Astronomy, 2, 102
\bibitem[MacLeod et al. (2019)]{mac19} MacLeod, C. L., Green, P. J., Anderson, S. F., et al. 2019, \apj, 874, 8
\bibitem[MacLeod et al. (2010)]{mac10} MacLeod, C. L., Ivezic, Z., Kochanek, C. S., et al. 2010, \apj, 721, 1014 
\bibitem[MacLeod et al. (2016)]{mac16} MacLeod, C. L., Ross, N. P., Lawrence, A., et al. 2016, \mnras, 457, 389
\bibitem[Mainzer et al. (2014)]{mai14} Mainzer, A., Bauer, J., Cutri, R. M., et al. 2014, \apj, 792, 30
\bibitem[Marin et al. (2019)]{mar19} Marin, F., Hutsemekers, D., \& Agis Gonzalez, B. 2019, in Proceedings of the 2019 Annual Conference of the SF2A, arXiv:1909.02801
% \bibitem[Mathur et al. (2018)]{mat18} Mathur, S., Denney, K. D., Gupta, A., et al. 2018, \apj, 886, 123
\bibitem[Massey et al. (1988)]{mas88} Massey, P., Strobel, K., Barnes, J. V., et al. 1988, \apj, 328, 315
\bibitem[McElroy et al. (2016)]{mc16} McElroy, R. E., Husemann, B., Croom, S. M., et al. 2016, \aap, 593, L8
\bibitem[Miller \& Stone (1993)]{mis93} Miller, J. S., \& Stone, R. P. S. 1993, Lick Obs. Tech. Rep. 66 (Santa Cruz, CA: Lick Observatory) 
\bibitem[Nicastro (2000)]{nic00} Nicastro, F. 2000, \apjl, 530, 65
\bibitem[Osterbrock \& Ferland (2006)]{osf06} Osterbrock, D. E., \& Ferland, G. J. 2006, Astrophysics of Gaseous Nebulae and Active Galactic Nuclei (2nd ed.; Sausalito, CA: University Science Books)
\bibitem[Pan et al. (2021)]{pan21} Pan, X., Li, S. -L., \& Cao, X. 2021, arXiv:astro-ph/2103:00828, accetped by \apj
\bibitem[Parker et al. (2016)]{par16} Parker, M. L., Komossa, S., Kollatschny, W., et al. 2016, \mnras, 461, 1927
% \bibitem[Raimundo et al. (2019)]{rai19} Raimundo, S. I., Vestergaard, M., Koay, J. Y., Lawther, D., Casasola, V., \& Peterson, B. M. 2019, \mnras, 486, 123
% \bibitem[Ross et al. (2018)]{ros18} Ross, N. P., Ford, K. E. S., Graham, M., et al. 2018, \mnras, 480, 4468 
\bibitem[Peterson et al. (2000)]{pet00} Peterson, B. M., McHardy, I. M., Wilkes, B. J., et al. 2000, \apj, 542, 161
\bibitem[Popovic et al. (2004)]{pop04} Popovic, L. C., Mediavilla, E., Bon, E., \& Ilic, D. 2004, \aap, 423, 909
\bibitem[Popovic et al. (2003)]{pop03} Popovic, L. C., Mediavilla, E. G., Bon, E., Stanic, N., \& Kubicela, A. 2003, \apj, 599, 185 
\bibitem[Risaliti et al. (2009)]{ris09} Risaliti, G., Young, M., \& Elvis, M. 2009, \apjl, 700, 6 
\bibitem[Ruan et al. (2016)]{rua16} Ruan, J. J., Anderson, S. F., Cales, S. L., et al. 2016, \apj, 826, 188
\bibitem[Ruan et al. (2019)]{rua19} Ruan, J. J., Anderson, S. F., Eracleous, M., Green, P. J., Haggard, D., MacLeod, C. L.\, Runnoe, J. C., \& Sobolewska, M. A. 2019, \apj, 883, 76 
% \bibitem[Rumbaugh et al. (2018)]{rum18} Rumbaugh, N., Shen, Y., Morganson, E., et al. 2018, \apj, 854, 160
\bibitem[Runnoe et al. (2016)]{tun16} Runnoe, J. C., Cales, S., Ruan, J. J., et al. 2016, \mnras, 455, 1691
%\bibitem[Schlegel et al. (1998)]{sch98} Schlegel, D. J., Finkbeiner, D. P., \& Davis, M. 1998, \apj, 500, 525
\bibitem[Schlafly \& Finkbeiner (2011)]{scf11} Schlafly, E. F., \& Finkbeiner, D. P. 2011, \apj, 737, 103
\bibitem[Shapovalova et al. (2010)]{sha10} Shapovalova, A. I., Popovic, L. C., Burenkov, A. N., et al. 2010, \aap, 509, 106
\bibitem[Shappee et al. (2014]{sha14} Shappee, B. J., Prieto, J. L., Grupe, D., et al. 2014, \apj, 788, 48
\bibitem[Sheng et al. (2017)]{she17} Sheng, Z., Wang, T., Jiang, N., et al. 2017, \apjl, 846, 7
\bibitem[Sheng et al. (2020)]{she20} Sheng, Z., Wang, T., Jiang, N., et al. 2020, \apj, 889, 46  
\bibitem[Sniegowska et al. (2020)]{sni00} Sniegowska, M., Czerny, B., Bon, E., \& Bon, N. 2020, \aap, 641, 167  
\bibitem[Stern et al. (2018)]{ste18} Stern, D., McKernan, B., Graham, M. J., et al. 2018, \apj, 864, 27
\bibitem[Storchi-Bergmann et al. (2017)]{sto17} Storchi-Bergmann, T., Schimoia, J. S., Peterson, B. M., Elvis, M., Denney, K. D., Eracleous, M., \& Nemmen, R. S. 2017, \apj, 835, 236
\bibitem[Strateva et al. (2003))]{atr03} Strateva, I. V., Strauss, M. A., Hao, L., et al. 2003, \aj, 126, 1720
\bibitem[Szalai et al. (2019)]{sza19} Szalai, T., Zsiros, S., Fox, O. D., Pejcha, O., \& Muller, T. 2019, \apjs, 241, 38 
\bibitem[Tody (1986)]{tod86} Tody, D. 1986, Proc. SPIE, 627, 733
\bibitem[Tody (1992)]{tod92} Tody, D. 1992, in ASP Conf. Ser. 52, Astronomical Data Analysis Software
and Systems II, ed. R. J. Hanisch, R. J. V. Brissenden, \& J. Barnes (San Fransisco, CA: ASP), 173
\bibitem[Trakhtenbrot et al. (2019)]{tra19} Trakhtenbrot, B., Arcavi, I., MacLeod, C. L., et al. 2019, \apj, 883, 94
\bibitem[Veron et al. (2002)]{ver02} Veron, P., Goncalves, A. C., \& Veron-Cetty, M. -P. 2002, \aap, 384, 862
\bibitem[Veron-Cetty \& Veron (2006)]{vev06} Veron-Cetty, M.-P., \& Veron, P. 2006, \aap, 455, 773
\bibitem[Wang et al. (2011)]{wan11} Wang, J., Mao, Y. F., \& Wei, J. Y. 2011, \apj, 741, 50
\bibitem[Wang et al. (2020a)]{wan20a} Wang, J., Xu, D. W., Sun, S. S., Feng, Q. C. Li T. R., Xiao, P. F., \& Wei, J. Y. 2020a, \aj, 159, 245
\bibitem[Wang et al. (2019)]{wan19} Wang, J., Xu, D. W., Wang, Y., Zhang, J. B., Zheng, J., \& Wei, J. Y. 2019, \apj, 887, 15
\bibitem[Wang et al. (2018a)]{wan18a} Wang, J., Xu, D. W., \& Wei, J. Y. 2018a, \apj, 852, 26
\bibitem[Wang et al. (2018b)]{wan18b} Wang, J., Xu, D. W., \& Wei, J. Y. 2018b, \apj, 858, 49 
\bibitem[Wang et al. (2020b)]{wan20b} Wang, J., Xu, D. W., \& Wei, J. Y. 2020b, \apj, 901, 1
\bibitem[Wang \& Bon (2020)]{wab20} Wang, J. -M., \& Bon, E. 2020, \apjl, 643, 9
\bibitem[Woo et al. (2017)]{woo17} Woo, J. H., Son, D., \& Bae, H. J. 2017, \apj, 839, 120
\bibitem[Wright et al. (2010)]{wri10} Wright, E. L., Eisenhardt, P. R. M., Mainzer, A. K., et al. 2010, \aj, 140,
1868
\bibitem[XMM-SSC (2018)]{xmm18} XMM-SSC, 2018, VizieR Online Data Catalog: XMM-Newton slew survey Source Catalogue, version 2.0, {https://ui.adsabs.harvard.edu/abs/2018yCat.9053....0X}
\bibitem[Yan et al. (2019)]{yan19} Yan, L., Wang, T. G., Jiang, N., et al. 2019, \apj, 874, 44
\bibitem[Yang et al. (2018)]{yan18} Yang, Q., Wu, X. B., Fan, X. H., et al. 2018, \apj, 862, 109
\bibitem[Zhang et al. (2013)]{zha13} Zhang, K., Wang, T., Gaskell, C. M., \& Dong, X. 2013, \apj, 762, 51

\end{thebibliography}
\end{document}